\documentclass[a4paper,12pt]{article}
\usepackage{tikz}
\usepackage{graphicx}
\usepackage{yfonts}
\usepackage{mathrsfs}
\usetikzlibrary{trees}
\usetikzlibrary{positioning,arrows}
\usetikzlibrary{decorations.pathmorphing}
\usetikzlibrary{decorations.markings}
\usetikzlibrary{decorations.text}
\usetikzlibrary{fit,chains,calc,shapes.geometric,intersections}
\usepackage[none]{hyphenat}

\pdfoutput=1
\usepackage{color}
\usepackage{amssymb}
\usepackage{epsfig}

\usepackage{graphicx}

\def\L{\mathcal L}
\def\e{\varepsilon}

\newcommand{\wt}{\widetilde}

\usepackage{amsmath}
\usepackage{amsfonts}

\usepackage{tikz}
\usetikzlibrary{matrix,positioning,decorations.pathreplacing}

\begin{document}

\tikzset{
particle/.style={thick,draw=black, postaction={decorate},
    decoration={markings,mark=at position #1 with {\arrow[black]{triangle 45}}}},
boson/.style={decorate, draw=black, thick,
    decoration={coil,aspect=0}}
 }

\def\a{\alpha}
\def\b{\beta}
\def\c{\chi}
\def\d{\delta}
\def\e{\epsilon}
\def\f{\phi}
\def\g{\gamma}
\def\h{\eta}
\def\i{\iota}
\def\j{\psi}
\def\k{\kappa}
\def\la{\lambda}
\def\m{\mu}
\def\n{\nu}
\def\o{\omega}
\def\p{\pi}
\def\q{\theta}
\def\r{\rho}
\def\s{\sigma}
\def\t{\tau}
\def\u{\upsilon}
\def\x{\xi}
\def\z{\zeta}
\def\D{\Delta}
\def\F{\Phi}
\def\G{\Gamma}
\def\J{\Psi}
\def\L{\Lambda}
\def\O{\Omega}
\def\P{\Pi}
\def\Q{\Theta}
\def\S{\Sigma}
\def\U{\Upsilon}
\def\X{\Xi}

\def\ve{\varepsilon}
\def\vf{\varphi}
\def\vr{\varrho}
\def\vs{\varsigma}
\def\vq{\vartheta}

\def\dg{\dagger}                                     
\def\ddg{\ddagger}                                   
\def\wt#1{\widetilde{#1}}                    
\def\mt{\widetilde{m}_1}
\def\mti{\widetilde{m}_i}
\def\rt{\widetilde{r}_1}
\def\mtt{\widetilde{m}_2}
\def\mttt{\widetilde{m}_3}
\def\rtt{\widetilde{r}_2}
\def\mb{\overline{m}}
\def\VEV#1{\left\langle #1\right\rangle}        
\def\be{\begin{equation}}
\def\ee{\end{equation}}
\def\ds{\displaystyle}
\def\ra{\rightarrow}

\def\bea{\begin{eqnarray}}
\def\eea{\end{eqnarray}}
\def\NO{\nonumber}
\def\Bar#1{\overline{#1}}


\def\pl#1#2#3{Phys.~Lett.~{\bf B {#1}} ({#2}) #3}
\def\np#1#2#3{Nucl.~Phys.~{\bf B {#1}} ({#2}) #3}
\def\prl#1#2#3{Phys.~Rev.~Lett.~{\bf #1} ({#2}) #3}
\def\pr#1#2#3{Phys.~Rev.~{\bf D {#1}} ({#2}) #3}
\def\zp#1#2#3{Z.~Phys.~{\bf C {#1}} ({#2}) #3}
\def\cqg#1#2#3{Class.~and Quantum Grav.~{\bf {#1}} ({#2}) #3}
\def\cmp#1#2#3{Commun.~Math.~Phys.~{\bf {#1}} ({#2}) #3}
\def\jmp#1#2#3{J.~Math.~Phys.~{\bf {#1}} ({#2}) #3}
\def\ap#1#2#3{Ann.~of Phys.~{\bf {#1}} ({#2}) #3}
\def\prep#1#2#3{Phys.~Rep.~{\bf {#1}C} ({#2}) #3}
\def\ptp#1#2#3{Progr.~Theor.~Phys.~{\bf {#1}} ({#2}) #3}
\def\ijmp#1#2#3{Int.~J.~Mod.~Phys.~{\bf A {#1}} ({#2}) #3}
\def\mpl#1#2#3{Mod.~Phys.~Lett.~{\bf A {#1}} ({#2}) #3}
\def\nc#1#2#3{Nuovo Cim.~{\bf {#1}} ({#2}) #3}
\def\ibid#1#2#3{{\it ibid.}~{\bf {#1}} ({#2}) #3}

\title{
\vspace*{1mm}
\bf Probing relic neutrino radiative decays with 21 cm cosmology}
\author{
{\Large M.~Chianese$^1$, P.~Di Bari$^1$, K.~Farrag$^{1,2}$ and R.~Samanta$^1$}
\\
$^1${\it Physics and Astronomy}, 
{\it University of Southampton,} \\
{\it  Southampton, SO17 1BJ, U.K.} \\
$^2${\it School of Physics and Astronomy}, 
{\it Queen Mary, University of London} \\
{\it  London, E1 4NS, U.K.} \\
}

\maketitle \thispagestyle{empty}

\vspace{-10mm}

\begin{abstract}
We show how 21 cm cosmology can test  relic neutrino radiative decays into sterile neutrinos. 
Using recent EDGES results,  we derive constraints on the  lifetime of the decaying neutrinos. 
If the EDGES anomaly will be confirmed, then  there are two solutions, one for much longer and one for much shorter lifetimes than the
age of the universe, showing  how relic neutrino radiative decays can explain the anomaly in a simple way.
We also show how to combine EDGES results with those from radio background observations, showing that potentially the
ARCADE 2 excess can be also reproduced together with the EDGES anomaly within the proposed non-standard cosmological scenario. 
Our calculation of the specific intensity at the redshifts probed by EDGES can be 
also applied  to the case of decaying dark matter 
and it also corrects a flawed expression used  in previous literature. 
\end{abstract}

\section{Introduction}

With 21 cm cosmology we are entering a new exciting phase in the study of the history of the universe  
and how this can be used to probe fundamental physics  \cite{review1,review2}.  
Observations of the redshifted $21\,{\rm cm}$ line
of neutral hydrogen\footnote{We refer to hydrogen-1 (protium). Deuterium has an analogous line but at 92 cm.}  
from the emission or absorption of the cosmic microwave background radiation (CMB)
by the intergalactic medium, can test the cosmic history at redshifts $z \sim 5$---$1100$.\footnote{We
 will discuss this range in more detail in Section 2.}
This range corresponds to those three  periods, after recombination,  on which we have fragmentary information:
 the {\em dark ages}, from recombination at $z_{\rm rec} \simeq 1100$ to $z \simeq 30$, when first
 astrophysical sources start to form; 
 the {\em cosmic dawn}, from $z \simeq 30$ to  the time when reionisation begins at $z \simeq 15$;
 the  {\em Epoch of Reionisation} (EoR), from $z\simeq 15$  to $z \simeq 6.5$
 when reionisation ends.\footnote{The redshift boundaries of these stages are, in fact, very model dependent. 
 Definite values strongly depend on astrophysical parameters (e.g., see \cite{cohen2017} for a recent parameter study). 
 The approximate value $z\simeq 6.5$ for the end of EoR corresponds to a particular
 model shown in Fig.~1 in \cite{review1} and references therein.} 
 In this way observations of the cosmological 21 cm line global signal 
can test the standard $\L$CDM cosmological model during a poorly period of the cosmic history, 
considering that the  most distant galaxy is located at $z=11.1$ \cite{GN-z11}.

Intriguingly, the EDGES ({\em Experiment to Detect the Global Epoch of Reionisation Signature}) collaboration
claims to have discovered an absorption signal in the CMB radiation spectrum corresponding
to the redshifted 21 cm line at $z \simeq 17.2$ with an amplitude  about twice  the expected value \cite{EDGES}.
This represents a $\sim 3.8\sigma$ deviation from the predictions of the $\L$CDM model and
for this reason the EDGES anomaly has drawn great attention. 
It should be said that  another group \cite{concerns}, re-analysing publicly available  EDGES 
data and using exactly their procedures, finds almost identical results
but they claim that `the fits imply either non-physical properties for the ionosphere 
or unexpected structure in the spectrum of foreground emission (or both)' concluding that their results
`call into question the interpretation of these data as an unambiguous detection of the cosmological 21-cm absorption signature.'
Therefore, more observations will  be necessary to confirm not only the anomaly but even the absorption signal. 

In the light of these recent experimental developments, it is anyway interesting to think of possible non-standard cosmological scenarios 
that can be tested with 21 cm signal observations at high redshifts
and that might either explain the EDGES anomaly (if confirmed) or in any case be constrained. 
The EDGES anomaly can be expressed in terms of a  value of the
photon-to-spin temperature ratio $T_{\g}(z)/T_S(z)$ at redshifts $z=15$---$20$,
where the absorption profile is observed, that is about twice what is expected in a standard cosmological scenario.
This can be  of course due either to a larger value of $T_\g(z)$ or a smaller value of $T_S(z)$ or some combination of the two. 
In this Letter, we  show how radiative decays of the lightest 
relic neutrinos can explain the EDGES anomaly producing, after recombination, 
a non-thermal early photon background able to rise $T_\g(z)$ above the CMB value.
A similar scenario, recently revisited in \cite{diego}, where heavier relic neutrinos
decay radiatively into lighter ordinary neutrinos \cite{berezhiani,rt,serpico} is ruled out since it
requires degenerate neutrino masses now excluded by the {\em Planck} upper bound
$\sum_i m_i \lesssim 0.17\,{\rm eV}\,(95\% \, {\rm C.L.})$ \cite{planck16} and since, 
it requires a too large effective magnetic moment responsible for the decay.
In our scenario the lightest relic neutrinos decay radiatively into 
sterile neutrinos and this allows to circumvent both bounds.\footnote{
See end of Section 3 for more details on this point while for an updated
review on neutrino mass constraints see for example \cite{PDG}.}

The paper is organised as follows.  
In Section 2 we briefly review 21 cm cosmology and the EDGES results. 
In Section 3 we discuss how lightest relic neutrinos radiative decays can explain the EDGES anomaly.
Finally, in Section 4, we draw the conclusions.

\section{21 cm cosmology and EDGES results}

The 21 cm line is associated with the hyperfine energy splitting between the two energy levels of the  1s ground state of  the
hydrogen atom characterised by a different relative orientation of electron and proton spins: anti-parallel  for the singlet level
with lower energy, parallel for the triplet level with higher energy. The energy gap between the two levels
and, therefore, of the absorbed or emitted photons at rest,
 is $E_{21} = 5.87\,\m{\rm eV}$ corresponding to a 21 cm line rest frequency 
 $\nu_{21}^{\rm rest} = 1420\,{\rm MHz}$. 

A shell of neutral hydrogen at a given redshift $z \lesssim z_{\rm rec}$, after recombination,
can then act, thanks to the 21 cm transitions, as a detector of the  background photons produced at higher redshifts.
In standard cosmology this background is just given by the CMB thermal radiation with temperature 
$T_{CMB}(z) = T_{CMB,0}\, (1+z)$, where $T_{CMB,0} = 2.725\,K \simeq 2.35 \times 10^{-4}\,{\rm eV}$.  

This possibility relies on the {\em brightness contrast} between the intensity of the 21 cm signal from the 
shell of neutral hydrogen gas at redshift $z$ and the background radiation 
at the observed (redshifted) frequency $\nu_{21}(z)=\nu_{21}^{\rm rest}/(1+z)$. 
The brightness contrast can be expressed in terms of the {\em 21 cm brightness temperature} (relative to the 
photon background) \cite{zaldarriaga}:
\be\label{T21}
T_{21}(z) \simeq  23\,{\rm mK} \, (1+\d_B)\, x_{H_I}(z) \,\left({\O_B\,h^2 \over 0.02}\right)\,
\left[\left({0.15 \over \O_{\rm m}\,h^2}\right)\,
\left({1+z \over 10} \right)  \right]^{1/2}  \,\left[
1 - {T_\g(z) \over T_S(z)} \right] ,
\ee
where $\O_B\,h^2 = 0.02226$ and $\O_{\rm m}\,h^2=0.1415 $ \cite{planck15} are respectively the baryon and matter abundances,  
 $\d_B$ is the baryon overdensity, $x_{H_I}$ is the fraction of neutral hydrogen, $T_{\gamma}(z)$ is the effective temperature, 
at frequency $\nu_{21}(z)$, of the photon background radiation (coinciding with 
$T_{CMB}(z)$ in standard cosmology) and $T_S(z)$  is the {\em spin temperature}  parameterising
the ratio of the population of the excited state $n_1$ to that one of the ground state $n_0$ in such a way that
\be
{n_1 \over n_0}(z) \equiv {g_1 \over g_0}  e^{-{E_{21} \over T_S(z)}} \,  \,  ,
\ee
where $g_1/g_0 =3$ is the ratio of the statistical degeneracy factors of the two levels. 
Clearly if $x_{H_I}$ vanishes, then there is no signal, since in that
case all hydrogen would be reionised and there cannot be any 21 cm transition. 
The spin temperature is related to $T_{\rm gas}$, the kinetic temperature of the gas,  by\footnote{This is an approximated 
relation valid for $T_{\rm gas} \simeq T_c$, where $T_c$
is the colour temperature parameterising the intensity of the UV radiation emitted by 
the astrophysical sources. Notice that out of the four different temperatures we introduced, only
$T_{CMB}$ and $T_{\rm gas}$ are genuine thermodynamic temperatures associated to a thermal
distribution.}
\be
\left(1- {T_{\gamma} \over T_S}\right) \simeq {x_c + x_\a \over 1 + x_c + x_\a }\, \left(1- {T_{\gamma} \over T_{\rm gas}}\right)   \,   , 
\ee
where $x_\a$ and $x_c$ are coefficients describing the coupling between 
the hyperfine levels and the gas.  In the limit of strong coupling, for $x_\a + x_c \gg 1$,
one has $T_S = T_{\rm gas}$, while in the limit of no coupling, for $x_\a = x_c =0$, one has $T_S = T_{\gamma}$
and in this case there is no signal.  The evolution of $T_{21}$ with redshift can be schematically described by five main stages \cite{review1,review2}: 
\begin{itemize}
\item[(i)]  In a first stage  after recombination,  during the dark ages,
the gas is still coupled to radiation thanks to a small but non negligible
amount of free electrons that still interacts via Thomson scatterings with the photon background. In this case one has
$T_{\g} = T_{\rm gas} = T_S$ and consequently $T_{21} =0$, i.e., there is no signal.\footnote{This conclusion is 
approximate and a very small signal is present even at high redshifts mainly due to the fact that
$T_c$ deviates slightly from $T_{\rm gas}$. 
This has been  studied recently in detail in \cite{breysse} and it was found
$-T_{21} \simeq 2.5\,{\rm mK}$ at $z \simeq 500$.}
This stage lasts until
the gas starts decoupling from radiation above $z_{\rm dec}^{\rm gas} \simeq 150$. At this time the gas temperature cools down more
rapidly than CMB radiation,  with $T_{\rm gas}(z) = T(z_{\rm dec})\,(1+z_{\rm dec})^2$. 
\item[(ii)] In  a second stage, approximately for $250 \gtrsim z \gtrsim 30$, still during the dark ages and
with the precise boundary values depending on different cosmological details,  one has approximately $T_S \simeq T_{\rm gas}$, since
gas collisions are efficient enough to couple $T_S$ to $T_{\rm gas}$. In this case one has $T_{21} < 0$ 
and an early absorption signal is expected.
\item[(iii)] At $z\simeq 30$ the gas becomes so rarified 
that the collision rate becomes too low to enforce $T_S \simeq T_{\rm gas}$ and in this case one enters a regime
where  $x_a + x_c \ll 1$ and $T_S \simeq T_{\gamma}$. In this stage, during the cosmic dawn,
one has $T_{21} \simeq 0$ and again the 21 cm
global signal is suppressed.\footnote{A detailed description and in particular how suppressed the signal is
in this stage depends on various astrophysical parameters \cite{cohen2017}.}

  \item[(iv)] At $z\simeq 30$, gas also starts collapsing under the action of dark matter and first astrophysical sources start
  to form with emission of Ly$\a$ radiation that is able, through Wouthuysen-Field effect \cite{WF}, 
  to gradually couple again $T_S$ to $T_{\rm gas}$. 
  In the redshift range $z_h \lesssim z \lesssim 25$, where $z_h \simeq 10$---$20$ 
  is the redshift at the heating transition\cite{review1} (this stage starts during the cosmic dawn  and can
  last until the epoch of reionization has begun at $z \simeq 15$), 
  one can again have $T_{21} < 0$, implying an absorption signal. 
  This is within the range tested by EDGES whose results seem to confirm the existence of the absorption signal.
\item[(v)]  In a fifth stage, for $z \lesssim z_h$ (depending on the precise value of $z_h$ this stage
can either start during cosmic dawn and ending during the epoch of reionization 
or entirely occurring during the latter), the gas gets reheated by the astrophysical radiation and 
$T_S \simeq T_{\rm gas} > T_{\gamma}$, so that $T_{21}$ turns positive and one has an emission signal from the regions that are not fully ionised.
Eventually all gas gets ionised until the fraction of neutral hydrogen vanishes and  the signal switches off again.\footnote{In this stage
the signal crucially depends on astrophysics and it should be said that not in all scenarios $T_{\rm gas}$ becomes larger than
$T_{\gamma}$ and in this case the emission signal is missing \cite{fialkov}.}
\end{itemize}


EDGES High and Low band antennas probe the frequency ranges 90-200 MHz and 50-100 MHz 
respectively overall measuring the 21-cm signal from between redshift 6 and 27, 
which corresponds to an age of the universe between 100 Myr and 1 Gyr 
and includes the epochs of reionization and cosmic dawn,
when first astrophysical sources form and a second stage of absorption signal is predicted
(the fourth and fifth stage in the description above).
The EDGES collaboration found an absorption profile approximately in the range $z=15$---$20$ with the minimum 
at $z_{\rm E} \simeq 17.2$, corresponding to $\nu_{21}(z_{\rm E})\simeq 78\,{\rm MHz}$,
with a 21 cm brightness temperature $T_{21}(z_{\rm E})=-0.5_{-0.5}^{+0.2}\,{\rm K}$
at $99\%\,{\rm C.L.}$, including estimates of systematic uncertainties. 
From Eq.~(\ref{T21}), since $(1+\d_{\rm b})\,x_{H_I}(z_{\rm E}) \simeq 1$, this translates into $T_{\g}(z_{\rm E})/T_S(z_{\rm E}) =15^{+15}_{-5.5}$.
On the other hand, at the centre of the absorption profile detected by EDGES, one expects, assuming the $\L$CDM model,
 $T_{\g}(z_{\rm E}) = T_{CMB}(z_{\rm E}) = T_{CMB,0}\,(1+z_{\rm E}) \simeq 50\,{\rm K}$
and $T_{\rm gas}(z_{\rm E}) \simeq T_{CMB}(z_{\rm dec}^{\rm gas})\,(1+z_{\rm E})^2/(1+z_{\rm dec}^{\rm gas})^2 \simeq 7\,{\rm K}$,
where we indicated with $z_{\rm dec}^{\rm gas}\simeq 150$ and $T_{\rm dec}^{\rm gas} \simeq 410\,{\rm K}$ respectively 
the  redshift and the temperature at the time when the gas decoupled from radiation.  From  Eq.~(\ref{T21}) one then 
immediately finds $T_{21}(z_{\rm E}) \gtrsim  -0.2\,{\rm K}$, where the minimum is saturated for $T_{S}(z_{\rm E}) = 
T_{\rm gas}(z_{\rm E})$ and corresponds to $T_{\g}(z_{\rm E})/T_{\rm gas}(z_{\rm E}) \simeq 7$.
Therefore, the best fit value for $T_{21}(z_E)$ is about $2.5$ lower than expected within the $\L$CDM.
Even at $99\,\%{\rm C.L.}$ it is still $50\%$ lower. 
 
If this anomalous result will be confirmed and astrophysical solutions ruled out, 
then, very interestingly, it can be regarded as the effect of some non-standard cosmological mechanism.
For example, it has been proposed that a (non-standard) interaction of the baryonic gas 
with the much colder dark matter component would cool down $T_{\rm gas}$, and consequently $T_S$, below the 
predicted $\L$CDM value \cite{barkana}. 
Another possibility is that $T_{\rm gas}$ is lower because the gas decouples earlier so that $z_{\rm dec}^{\rm gas} > 150$.
For example for $z_{\rm dec}^{\rm gas} \simeq 300$, one has $T_{\rm gas}(z_{\rm E}) \simeq 3.5\,{\rm K}$, i.e.,  halved compared to the
value predicted within the $\L$CDM, and this would reconcile the tension between $\L$CDM prediction and the EDGES result. 
Models of early dark energy have been proposed to this extent, but these are strongly ruled out by observations of 
the CMB temperature power spectrum \cite{hillbaxter}.
A third possibility is that some non-standard source  could produce a non-thermal additional
component of soft photons effectively increasing $T_{\gamma}$ above $T$ at frequencies around $\nu_{21}(z_{\rm E})$. 
For example, these could be produced by  dark matter annihilations and/or decays \cite{darkmatter,marzola} and also
give a signal at other frequencies for example addressing the ARCADE 2 excess at higher ($\sim {\rm GHz}$) frequencies
\cite{ARCADE2} that, however, has not been confirmed by another group using ATCA data \cite{ATCA}.
Conversion of dark photons  into soft photons has also been proposed as a solution to the EDGES anomaly \cite{darkphotons}.

In the next section we present a mechanism for the production of a non-thermal soft photon
component relying  on  relic neutrinos radiative decays into sterile neutrinos.  Even if the EDGES anomaly will not be confirmed, we show that 
the  EDGES results tighten the existing constraints \cite{rt,kt} on the parameters of the scenario.

\section{Relic neutrino radiative decays}

The 21 cm CMB photons  absorbed at $z_E$ fall clearly in the Rayleigh-Jeans tail 
since $E_{21} \ll T(z_{\rm E})$. In this regime the specific intensity 
depends linearly on temperature, explicitly 
\be
I_{CMB}(E,z) \equiv {d{\cal F}^{\g_{CMB}}_E \over dA\,dt \, dE \,d\O}  = {1 \over 4\pi} \, {d\varepsilon_{CMB} \over d E}
= {E^3 \over 4\,\pi^3}\,[e^{E/T_{CMB}}-1]^{-1} \, \stackrel{E \ll T_{CMB}}{\longrightarrow} \, {1 \over 4 \pi^3}\,T_{CMB}(z) \, E^2  \,  .
\ee
Only photons with energy $E_{21}$ at $z\simeq z_{\rm E}$ could be absorbed by the neutral hydrogen producing
a 21 cm absorption global signal.  The EDGES results can be explained by an additional non-thermal photon background with 
$I_{\rm nth}(E_{21}, z_{\rm E}) \simeq I_{CMB}(E_{21},z_{\rm E})$. 
The effective photon temperature $T_{\g}(E,z_E)$ at an arbitrary energy $E$ can  be simply calculated as
\be\label{Tgamma}
T_{\gamma}(E, z_E)=  {E \, \ln^{-1} \left(1+{E^3 \over 4\,\pi^3 \, I_\g(E,z_E)} \right)}  
\stackrel{E \ll T_\g}{\longrightarrow} {4 \pi^3 \over E^2} \, 
I_\g(E,z_E) \,  ,
\ee  
where we defined $I_\g(E,z_E)= I_{\rm nth}(E,z_E)+I_{CMB}(E,z_E)$.
The EDGES anomaly can be explained imposing 
$T_{\g}(E_{21},z_E)/T_{CMB}(z_E) = 2.15^{+2.15}_{-0.8}$,
or, equivalently, 
\be\label{R}
R \equiv {I_{\rm nth}(E_{21},z_E) \over I_{CMB}(E_{21},z_E)}=
{T_{\g_{\rm nth}}(E_{21},z_E) \over T_{CMB}(z_E)}= R_{E} \equiv 1.15^{+2.15}_{-0.8} \,   ,
\ee
where $T_{\g_{\rm nth}}$ is defined in terms of $I_{\rm nth}$ in the same way as $T_{\gamma}$ 
is defined in terms of $I_{\gamma}$ in Eq.~(\ref{Tgamma}). 
We consider the radiative decay of active neutrinos $\nu_i$  with mass $m_i$ and 
lifetime $\tau_i$ into a sterile neutrino $\nu_{s}$ with mass $m_{s}$, i.e.,  $\nu_i \ra \nu_{s} + \g$.
For definiteness we will refer to the case of  lightest neutrino decays corresponding to $i=1$. We will comment
at the end how our results simply change if one considers heavier neutrinos.  
If decays occur after matter-radiation decoupling, photons produced from the decays will not distort CMB
thermal spectrum but will give rise to a non-thermal $\g$ background \cite{rt} contributing to $R$.
For a given $m_1$ one has two limits for $m_s$: a quasi-degenerate limit for $m_1 \simeq m_s$ and
a limit  $m_s \ll m_1$.\footnote{The origin and properties of neutrino masses and mixing would be related to
extensions of the SM (e.g., grand-unified theories). Simplest models usually require the existence of very heavy
sterile neutrinos  ($m_{\rm s} \gg 100\,{\rm GeV}$) 
in the form of right-handed neutrinos. However, the existence of light sterile neutrinos 
cannot be excluded and many models have been proposed especially in connection with 
different neutrino mixing anomalies (for a review see for example \cite{giunti}).}

For $m_s \ll m_1$ the bulk of neutrinos, with $E \sim T_{CMB}$, necessarily decay when they are relativistic. This is easy to understand.
Let us introduce the scale factor $a = (1+z)^{-1}$ and its corresponding value $a_E \equiv (1+z_E)^{-1}$ at $z_E$.
In the matter-dominated regime we can  write $a(t) \simeq a_{E} \, \left(t / t_{E} \right)^{2 \over 3}$,
where $t_{\rm E} \simeq 222\, {\rm Myr}$ is the age of the universe at $z=z_E$ \cite{libro}.  
For neutrinos that decay at rest at time $t$ one has to impose $m_1  = 2\,E_{21} \, {a_{E} / a(t)}$
in order to have photons with the correct energy at $t_E$. 
Imposing that decays occur after recombination, otherwise the non-thermal
component would  thermalise or produce unacceptable distortions to the CMB spectrum,  
and of course before the time when photons are absorbed by neutral hydrogen, corresponding to
a condition $z_{\rm E} < z(t) < z_{\rm rec} \simeq 1100$, one finds
$0.012\,{\rm meV} \lesssim m_1 \lesssim 0.71\,{\rm meV}$,  
showing that the $\nu_1$'s are too light to be treated  non-relativistically for $m_s \ll m_1$.\footnote{This also 
shows that the two heavier neutrinos radiative decays would produce
photons at too high frequencies, considering that $m_2 \geq m_{\rm sol} \simeq 9\,{\rm meV}$
and $m_3 \geq m_{\rm atm} \simeq 50\,{\rm meV}$, where the lower bounds are saturated in the
normal hierarchical limit. One could consider radiative decays $\nu_{2,3} \ra \nu_1 + \gamma$
and  in  the quasi-degenerate limit  $m_1 \gtrsim 0.12\,{\rm eV}$ 
photons with the correct energy would be produced. However, the upper bound
$m_1 \lesssim 0.07\,{\rm eV}$ placed by the {\em Planck} collaboration 
now rules out this possibility \cite{planck16}. Moreover these processes
need values of the effective  magnetic moment ruled out
by current experimental upper bound \cite{serpico,diego}.}

On the other hand the non-relativistic case can be realised in the quasi-degenerate limit for $m_1 \simeq m_s$
since  one can then have $m_1 \gg  T_{\nu}(z) \simeq (4/11)^{1/3}\,T(z)$ at the time when they decay.  Indeed at $z = z_E$
one has $T_{\nu}(z_E) \simeq 3\,{\rm meV}$. Since the current upper bound on the sum
of neutrino masses implies $m_1 \lesssim 50\,{\rm meV}$, one can well have $m_1 \simeq m_s \gg 3\,{\rm meV}$.
This implies $50 \gtrsim m_1/{\rm meV} \gtrsim 10$,\footnote{The lower bound $m_1 \gtrsim 10\,{\rm meV}$ 
corresponds to $m_1 \gtrsim 3\,T_{\nu}(z_E)$
that is quite a conservative condition to enforce that the bulk of neutrinos are non-relativistic when they decay 
since remember that for a Maxwell-Boltzmann distribution $\sqrt{\langle v^2\rangle}=\sqrt{3\,T/m}$.
In this way the bulk of neutrinos have a kinetic energy that is negligible compared to the rest energy.} 
a window that will be fully tested by close future cosmological observations \cite{lesgourgues}.
Moreover in this (testable) non-relativistic and quasi-degenerate case
not only it is  easy to calculate $R$, as we will see, but also one obtains   the most conservative constraints  on 
$\tau_1$ and $\D m_1 \equiv m_1 - m_s$ since,
as we will point out, if neutrinos decay relativistically,  constraints get more stringent.

Let us then calculate $R$ for $m_1 \simeq m_s \gg T_{\nu}(z)$.  An emitted photon has an energy at decay $E = \D m_1 $.
Moreover let us suppose first that all neutrinos decay instantaneously at $t = \tau_1$ corresponding to a redshift
$z_{\rm decay}$ such that $a_{\rm decay} = (1+z_{\rm decay})^{-1} \simeq a_E\,(\tau_1/t_E)^{2/3}$.
 A sketchy representation of this toy model is given in Fig.~1. 
\begin{figure}
\begin{center}
\psfig{file=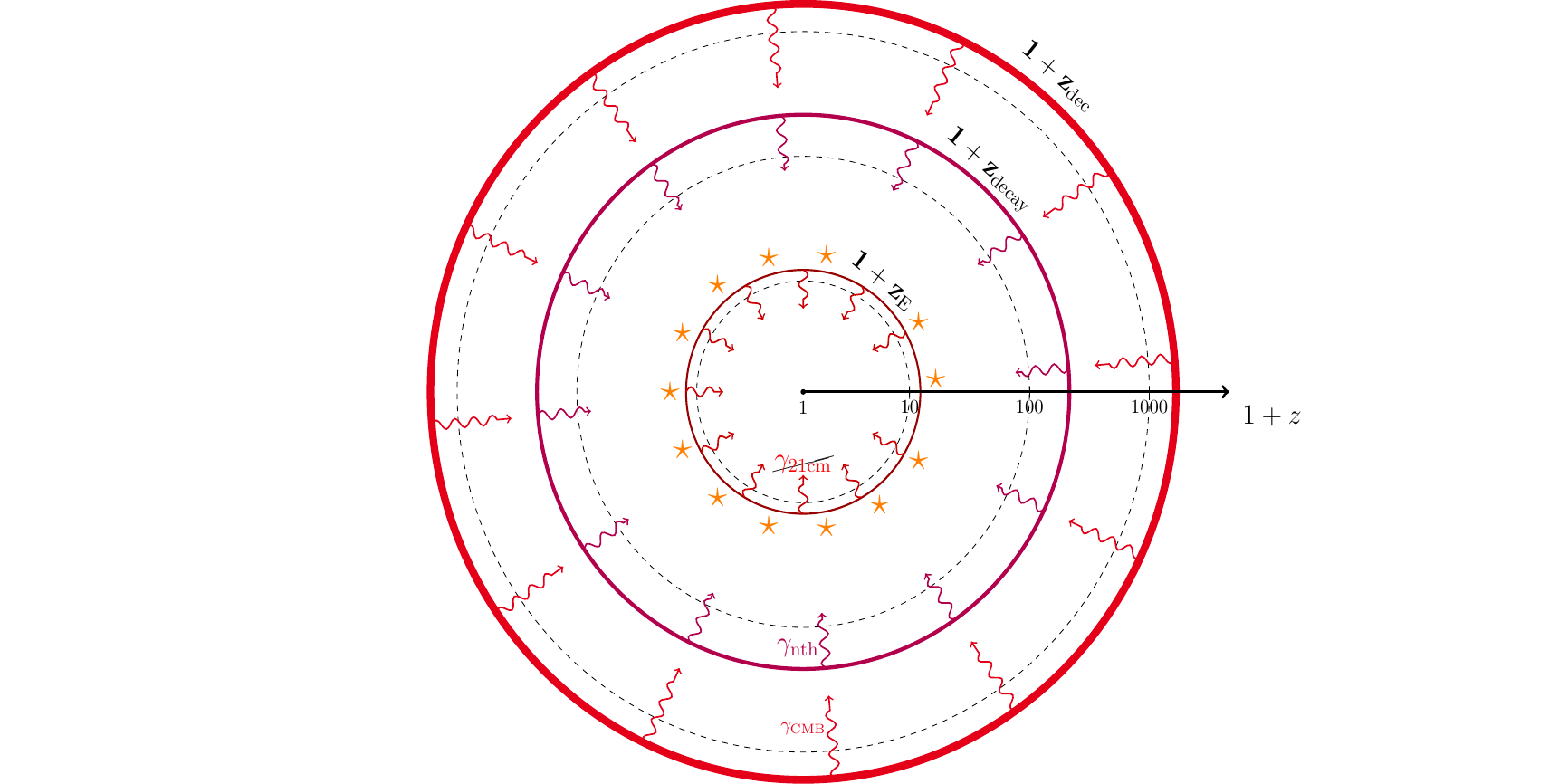,height=70mm,width=135mm} 
\end{center}
\vspace{-5mm}
\caption{Schematic picture describing the generation of the 21 cm absorption 
signal within a non-standard history of the universe  including $\nu_1$ radiative decays producing
a non-thermal photon background. 
At redshifts $z\sim z_{\text{E}}$ the first astrophysical sources couple the spin temperature
to the gas temperature (via Wouthuysen--Field effect) inducing the absorption of (redshifted) $\gamma_{21 {\rm cm}}$ photons.}
\label{fig1}
\end{figure}
Requiring that photons produced from neutrino decays are then (21 cm) absorbed at $z=z_E$, one has to impose
$\D m_1 = E_{21} \, a_E /a_{\rm decay}$, implying
an unrealistic fine-tuned relation between $\tau_1$, $E_{21}$ and $\D m_1$. In addition
it is easy to see that, since all photons produced in the decay contribute to the signal, 
one obtains a far too high value of $R$. This is because in this instantaneous decay description one has simply
\be
I_{\rm nth}(E_{21},z_{\rm E})= {n^{\infty}_{\nu_1}(z_{\rm E})  \over 4\pi} = 
 {3 \over 11} \, {\zeta(3) \over 2\,\pi^3}\,T_{CMB}^3(z_E) \,   ,
\ee
where $n^{\infty}_{\nu_1}(z_{\rm E})=(6/11)(\zeta(3)/\pi^2)\,T_{CMB}^3(z_E)$ is the relic neutrino number density at $z_E$ in the standard stable neutrino case. 
This gives straightforwardly:
\be\label{Rstar}
R \simeq R_{\star} \equiv
{6\,\zeta(3) \over 11} \,  \left[ {T_{CMB,0}\,(1+z_E) \over E_{21}}\right]^2  \simeq 3.5 \times 10^5 \,  ,
\ee
many orders of magnitude larger than $R_{E}$. 
However, this simplistic instantaneous description 
has the merit to show the natural normalisation for the specific intensity of the non-thermal photons in terms of  
$n^{\infty}_{\nu_1}(z_{\rm E})$ and for $R$ itself in terms of $R_{\star}$.

Let us now calculate  $R$  removing the instantaneous assumption. 
Writing the fluid equation for the energy density of non-thermal photons produced by $\nu_1$ decays \cite{rt,kt,warm},  
\be
{d\ve_{\g_{\rm nth}} \over dt} = {\D m_1 \over \tau_1}\, n_{\nu_1}^{\infty}(t)\,e^{-{t \over \tau_1}} - 
4\, \ve_{\g_{\rm nth}} \, H  \,  ,
\ee 
where $H \equiv \dot{a}/a$ is the expansion rate, one easily finds a solution in terms of a Euler integral
\be
\ve_{\g_{\rm nth}}(t_E)= {\D m_1 \over \tau_1}\, n_{\nu_1}^{\infty}(t_E)\, \int_0^{t_E}\,dt \, e^{-{t \over \tau_1}} \, {a(t) \over a(t_E)} \,  .
\ee
The integral is done over all times $t$ when photons are produced by neutrino decays with energy $\D m_1$
that is redshifted to an energy  $E(t,t_E) = \D m_1\,a(t)/a_E$ at $t_E$.
Photons with the correct energy $E_{21}$ at $t_E$
are produced at a specific time $t_{21}$ such that $a_{21}/a_E = E_{21}/\D m_1$,
where $a_{21} \equiv a(t_{21})$.  
Of course notice that imposing $z_{\rm rec} > z_{21} \equiv a_{21}^{-1} -1 > z_E$, one
would find  $E_{21} \lesssim \D m_1 \lesssim 0.35 \,{\rm meV}$, 
 analogously to the range found for $m_1$ in the case $m_s \ll m_1$. 
 However, since we are assuming that neutrinos are non-relativistic at decays, 
 and this implies $T_{\nu}(z_{21}) \simeq 0.18\,{\rm eV}\,(1+z_{21})/(1+z_{\rm dec})\lesssim m_1 \lesssim 50\,{\rm meV}$, 
 one finds $z_{21} \lesssim 275$, implying an even more restrictive range
 \be\label{range}
 E_{21} \lesssim \D m_1 \lesssim 0.9\times 10^{-4}\,{\rm eV} \,  .
 \ee
We can now easily switch from  time to  energy derivative finding
\be\label{Inth}
I_{\rm nth}(E_{21},z_E)  = {1 \over 4 \pi} \, {d\varepsilon_{\g_{\rm nth}} \over d E}
\, = {n_{\nu_1}^{\infty}(z_E) \over 4\,\pi} \,\left({E_{21}\over \D m_1}\right)^{3/2} \, 
{e^{-{t_E \over \t_1}\,\left({E_{21}\over \D m_1}\right)^{3/2}} \over H_E \, \tau_1}\,  ,
\ee
where $H_E \equiv H(t_E) \simeq 2/(3\,t_E)$.
From the definition of $R$ (see Eq.~(\ref{R})) and $R_{\star}$ (see Eq.~(\ref{Rstar})), one immediately obtains
\be
R = R_{\star}\, \left({E_{21}\over \D m_1}\right)^{3/2} \,{e^{-{t_E \over \t_1}\,\left({E_{21}\over \D m_1}\right)^{3/2}} \over H_E \, \tau_1} \,  .
\ee 
The condition $R \leq  R_{E}$, where the equality corresponds to the condition to explain the EDGES anomaly and the inequality 
implies constraints on $\tau_1$ and $\D m_1$,  can be put in the simple form\footnote{Of course 
one should also not forget that $\D m_1$ is constrained within the range 
in Eq.~(\ref{range}).}
\be
x\,e^{-x} = {2 \over 3}\,{R_{E} \over R_{\star}} = 2.2^{+4}_{-1} \times 10^{-6} \,  ,
\ee
where we defined
\be 
x \equiv {t_E \over \tau_1}\, \left({E_{21}\over \D m_1}\right)^{3/2} \,  .
\ee
There are clearly two solutions. A first one (referred to as `EDGES A' in Fig.~2) for $\tau_1 \gg t_E$ is  simply given by $x= 2.2^{+4}_{-1} \times 10^{-6}$,
from which one finds
\be\label{EDGESsol}
\tau_1 \geq 100_{-65}^{+ 300}\,t_0 \, \left({10^{-4}\,{\rm eV}\over \D m_1}\right)^{3/2}  \,  ,
\ee
where  $t_0 =13.8\,{\rm Gyr} = 4.35 \times 10^{17}\,{\rm s}$ is the age of the universe. 
\begin{figure}
\begin{center}
\psfig{file=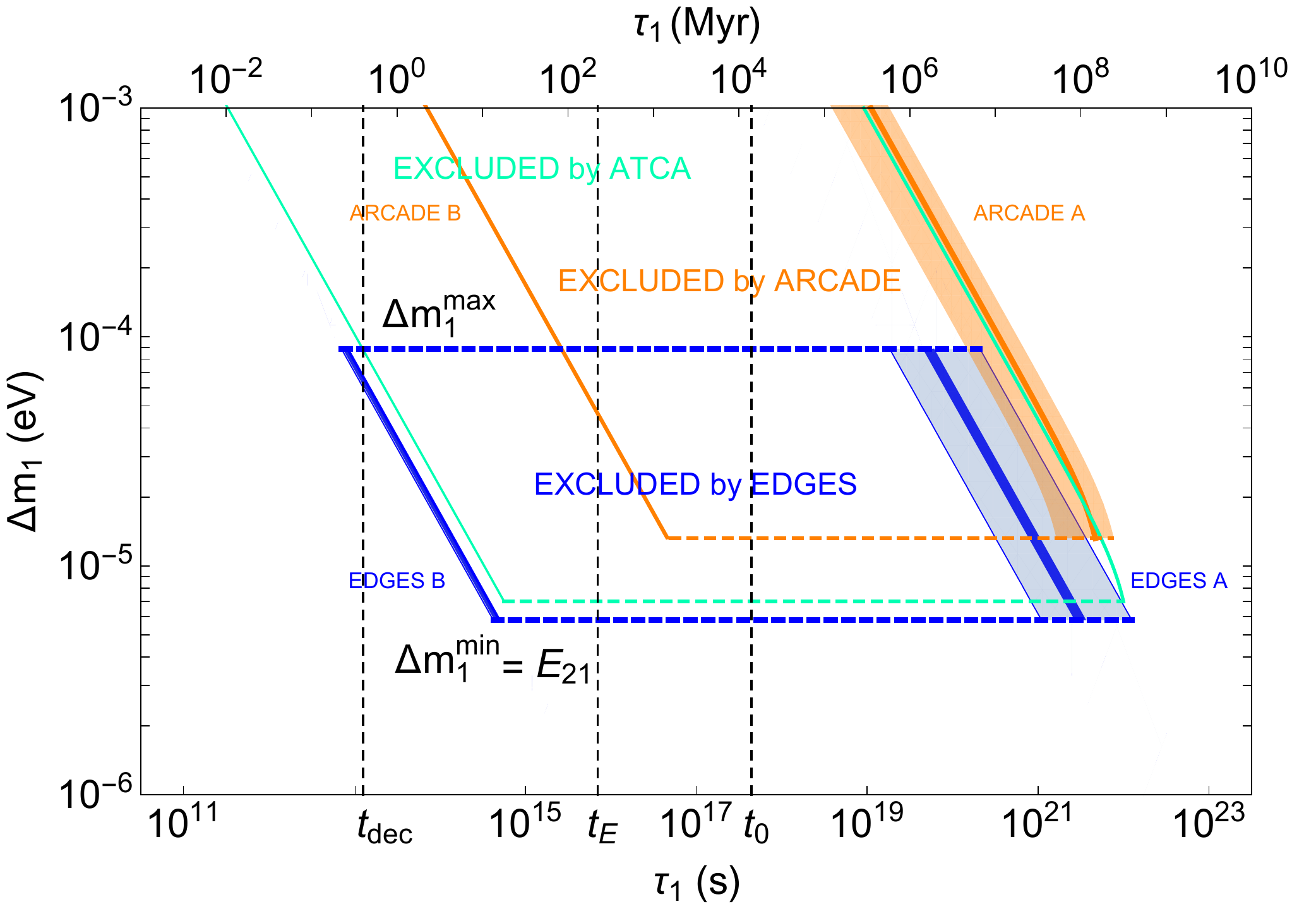,height=70mm,width=115mm} 
\end{center}
\vspace{-10mm}
\caption{The blue lines delimit the excluded region in the plane $\tau_1-\D m$ from EDGES data and the two solutions (A and B) addressing the 
EDGES anomaly. The orange area indicates the approximate solution to the ARCADE 2 excess.
The cyan area indicates the ATCA constraint.}
\label{fig2}
\end{figure}
A second solution (referred to as `EDGES B' in Fig.~2) is found for $\tau_1 \ll t_E$ and is given by $x =15.2^{+1.8}_{-0.5}$, from which one finds
\be
\tau_1 \leq 2.10^{+0.04}_{-0.24} \times 10^{5} \, {\rm yr} \, \left({10^{-4}\,{\rm eV}\over \D m_1}\right)^{3/2} \,  .
\ee
For this second solution one has to consider that decays should occur mainly after matter-radiation decoupling time in order to have 
a photon non-thermal background and one has to impose
 $\tau_1 \gtrsim t_{\rm dec} \simeq 3.71 \times 10^5 \,{\rm yr}$. Moroever, though photon energies
are much below the thermal bath temperature, they might produce too large deviations of CMB from a thermal spectrum. 
Even though this second solution is less appealing and likely not viable, it is also interesting  that 
one could in principle expect a number of neutrino species
at recombination lower than three if only a fraction of the decays are allowed to occur before recombination.

We have now also to consider  whether photons produced from neutrino decays might give visible
(wanted or unwanted) effects at other frequencies.  First of all one should worry of the CMB spectrum tested
by the the COBE-FIRAS instrument in the range of frequencies $(2$---$21)\,{\rm cm}^{-1}$
corresponding to $(60$---$600)\,{\rm GHz}$ or to energies ($0.25$---$2.5) \,{\rm meV}$ \cite{COBEFIRAS}.
However, since  $\D m_1 < 0.09\,{\rm meV}$ (see Eq.~(\ref{range})), in this non-relativistic 
scenario one completely circumvents
the constraints from CMB thermal spectrum measurements.

Radio background observations in the GHz frequencies can also test the scenario
either constraining it, as ATCA data do \cite{ATCA}, or even providing, with the ARCADE 2 excess \cite{ARCADE2}, another 
signal to be explained together with the EDGES anomaly.\footnote{Notice that 
in \cite{cowsik} the existence of the ARCADE 2 excess is questioned and it is proposed that a
a more realistic galactic model can reconcile measurements of uniform extragalactic brightness  
by ARCADE 2 with the expectations from known extragalactic radio source populations.}

 Let us see how they can be combined with 21 cm observations
to test relic lightest neutrino decays. In this case the results are given in terms of an effective temperature $T_{\rm rb}(E_{\rm rb})$ of the
radio background compared to the Rayleigh-Jeans tail of the CMB spectrum.  This time the detection of the produced photons
is made directly at the present time, while in the case of EDGES, as we discussed, photons produced by the decays are 
absorbed by the intergalactic medium at the time $t_E$. 
Therefore, now we have to impose
\be
I_{\rm nth}(E_{\rm rb},z=0) ={n^{\infty}_{\nu_1}(z=0)\over 4\,\pi}\,{e^{-{t(a_{\rm rb})\over \tau_1}} \over H(a_{\rm rb}) \, \tau_1}  = 
{1\over 4\,\pi^3}\,T_{\rm rb}\,E^2_{\rm rb}  \,  ,
\ee
where this time  $a_{rb} = E_{\rm rb}/\D m_1$. 
If we focus on the solution at $\tau_1 \gg t_0$, then the exponential can be neglected and 
using $H(a_{rb}) = \sqrt{\O_{M0}\,a_{rb}^{-3}+\O_{\L 0}}$
and defining $a_{\rm eq}^{M\L} \equiv (\O_{M0}/\O_{\L 0})^{1/3} \simeq 0.75$, one arrives to the
condition
\be\label{radio}
\tau_1 \geq {6\,\zeta(3) \over 11\,\sqrt{\O_{M0}}}\,{T_{CMB,0}^3 \, \left({E_{\rm rb}\over \D m_1}\right)^{3/2} \over T_{\rm rb} \, E^2_{\rm rb}} 
\left(1 + {a_{\rm rb}^3 \over a_{\rm eq}^{M\L}}\right)^{-1/2}
\, t_0  \,   , 
\ee
where again the equality holds in the case one wants to explain the ARCADE 2 excess  or the inequality in the case
one impose the constrain from the ATCA data.  The ARCADE 2 collaboration claims an excess with
$T_{\rm rb} = (62 \pm 10) \, {\rm mK}$ at a frequency $3.2\,{\rm GHz}$ corresponding to $E_{\rm rb} = 13.2 \times 10^{-6}\,{\rm eV}$
and from the condition (\ref{radio}) one finds
\be\label{ARCADE}
\tau_1 \simeq (800 \pm 200)\,\left( {10^{-4}\,{\rm eV} \over \D m_1} \right)^{3/2} \, 
\left(1 + {a_{\rm rb}^3 \over a_{\rm eq}^{M\L}}\right)^{-1/2}
\, t_0  \,  ,
\ee
a solution shown in Fig.~2 (`ARCADE A') in orange (at $99\%$ C.L.) together with the corresponding 
allowed range for $\D m_1$ found similarly to Eq.~(\ref{range}) with the
difference that now the energy at the production has to be redshifted at $z=0$ instead of $z=z_E$.
If this is compared with the condition we found to explain the EDGES anomaly Eq.~(\ref{EDGESsol}), one can see
that within the errors the two anomalies can be reconciled, in particular for the lowest values of $R_{E}$
corresponding to the highest  $\tau_1$ values, and of course with the help of $a_{rb}$ as close as possible to unity (its maximum value),
corresponding to $\D m_1 = E_{\rm rb}$. 
As in the case of the EDGES one can also consider a solution for $\tau_1 \ll t_0$ and in this case one finds, neglecting 
this time the small correction from $\Omega_{\L}$,
$\tau_1 \simeq (70 \pm 0.7)\,{\rm Myr}\, (10^{-4}\,{\rm eV}/ \D m_1)^{3/2}$. This is also show in Fig.~2 (`ARCADE B')  in
orange (at $99\%$ C.L.). However, this time it is clear that there is no overlap with the `EDGES B' solution and this somehow
makes even more remarkable that in the case of $\tau_1 \gg t_0$ we could find two overlapping solutions. 
One can also consider the ATCA constraints \cite{ATCA}
that place a ($3\,\s$) upper bound $T_{\rm rb} \lesssim 100\,{\rm mK}$ at a frequency of $1.75\,{\rm GHz}$.
In this case one finds the following ($99\%$) excluded region 
\be
3.3\times 10^5\,{\rm yr} \, \left({10^{-4}\,{\rm eV}\over \D m_1}\right)^{3/2} 
\lesssim \tau_1
\lesssim 660\,\left( {10^{-4}\,{\rm eV} \over \D m_1} \right)^{3/2} \, 
\left(1 + {a_{\rm rb}^3 \over a_{\rm eq}^{M\L}}\right)^{-1/2} \, t_0  \,   .
\ee
shown in Fig.~2 in light green.
Notice that this constraint does not apply to the narrow range
$ E_{21} < \D m_1 < E_{\rm rb} \simeq 7 \times 10^{-6}\,{\rm eV}$ 
(so that EDGES allows to extend the constraints at slightly lower values of $\D m_1$).\footnote{Recently
a study of the radio background data from the LWA1 Low Frequency Sky Survey (LLFSS)
at frequencies between $40\,{\rm MHz}$ and $80\,{\rm MHz}$ \cite{LWA}
has found an excess well described by a power law $T_{\rm rb} \simeq T_{\rm rb,0} \, (\nu/\nu_0)^{\beta}$ with $\nu_0=310\,{\rm MHz}$
and $\beta\simeq -2.58$, also fitting the ARCADE 2 results at much higher frequencies. 
For example at $\nu=80\,{\rm MHz}$ the survey finds $T_{\rm rb} = (1188 \pm 112)\,{\rm K}$.
This excess cannot be explained by our model since from Eq.~(\ref{radio}) one can see that it predicts
$T_{\rm rb} \propto E^{-0.5}$. If we fit the ARCADE 2 results, then we have a signal at $\nu = 80\,{\rm MHz}$, approximately 
the same frequency tested by EDGES,
that is about 100 times smaller than what LLFSS finds. Of course the LLFSS results do not 
exclude our model, they simply require an alternative explanation. More generally, they can be hardly 
reconciled with the EDGES anomaly  within a realistic model  since one would need a mechanism where the intensity
of the produced radiation increases by about $20$ times between $z \simeq z_E$ and today and this despite the
fact that the expansion dilutes a matter fluid number density, such as primordial black holes,  
by a factor $(1+z_E)^3$. Even if one finds a model where this huge enhancement of the intensity is realised, 
this has to be strongly fine-tuned to  match both results and this without considering the ARCADE 2 excess.}
Another interesting observation is that in the second stage of the evolution of $T_{21}$ that we outlined in Section 2,
for redshifts $250 \gtrsim z \gtrsim 30$, one expects an early absorption signal at $z\simeq 100$. 
If we extend the definition of $R$ at a generic redshift $z_{\rm absorption}$, one can easily see from our expressions that
this scales as $\propto \sqrt{1+z_{\rm absorption}}$. Therefore, the scenario predicts  a doubled
value of $R(z\simeq 100)$ compared to the one measured at $z_E$. This would be a powerful test of the scenario,
though consider that in order to have a signal also in the early absorption stage, 
the upper bound on $\D m_1$ in Eq.~(\ref{range}) 
gets more stringent: from the requirement  $z_{21} \lesssim 275$,  now  one obtains $\D m_1 \lesssim 3\,E_{21}$. 


The derivation of the constraints could be further extended going beyond the quasi-\\ degenerate limit $m_1 \simeq m_s$
implying necessarily going beyond the non-relativistic regime. In this case one has to take into account the thermal distribution function 
of neutrinos and from this derive the non-thermal distribution of photons solving a simple Boltzmann equation \cite{massotoldra}.
The factor $R$ gets reduced for fixed $\tau_1$ since the photon energy spectrum spreads at higher energies 
and at the energy $E_{21}$ at $z_E$ there are less photons and so the values of the  lifetime that are necessary
to explain the EDGES anomaly become shorter and this tends to generate 
a conflict with the constraints from radio observations and likely with the FIRAS-COBE data as well
since photon energies can be much higher.

Finally, let us comment that though we have considered for definiteness decays of the lightest neutrinos,
the results are also valid for heavier neutrinos of course with the replacement 
$(\tau_1,\D m_1) \ra (\tau_{2,3},\D m_{2,3})$. The only difference is that now they automatically respect the 
condition $m_{2,3} \gg 3\,{\rm meV}$ to be non-relativistic and of course in this case the lower bound $m_1 \gtrsim 10\,{\rm meV}$
does not hold so that the lightest neutrino mass can be arbitrarily small since lightest neutrinos do not play any role.

We should also say that of course, even though for definiteness we considered radiative decays into sterile neutrinos,
our results are valid for any other decay mode involving a light exotic particle. 
Our results can also be easily exported to the case of quasi-degenerate 
dark matter recently proposed in \cite{marzola} though notice that 
the correct way to calculate the specific intensity is Eq.~(\ref{Inth})
(replacing of course neutrino  with dark matter number density) 
that takes into account that only those photons produced before $t_E$ can be responsible for the signal
while the authors of \cite{marzola} incorrectly use an expression valid for photons detected at the present time. 
However notice that in the case of decaying dark matter the fact that 
the intensity of non-thermal photons has to be comparable to that of CMB photons, as required by EDGES,
is a coincidence. On the other hand, in the case of decays of active to light sterile neutrinos,
the abundance of relic active neutrinos is fixed by thermal equilibrium 
and this naturally produces a non-thermal photon intensity 
comparable to that of CMB photons.
One can think of a simple  model for example in terms of singular seesaw \cite{singular} extended with a
type II contribution \cite{singulartypeII}.  
In this case an active-sterile neutrino mixing is expected and one can have interesting
phenomenological consequences that can help testing the scenario.\footnote{Radiative decays would still 
generate an effective magnetic moment for active neutrinos but if the
mixing the sterile neutrino is sufficiently small, a condition easily realised 
especially for the A solution with very long lifetime, this can be well
below the upper bound from stellar cooling.}
For example, in addition
to obvious possible effects in neutrino oscillation experiments and in particular in solar neutrinos,
the fact that $m_s < m_1$ makes possible a mechanism of generation of a large lepton asymmetry
in the early universe \cite{footvolkas}  with possible testable effects in big bang nucleosynthesis 
and CMB acoustic peaks \cite{bbncmb}.\footnote{One could investigate whether such dynamical generation of
the asymmetry might suppress the thermalisation of a $\sim {\rm eV}$ sterile neutrino \cite{footvolkas}
required by the solution to the various anomalies \cite{maltoni}.}


\section{Conclusion}

We discussed a scenario where relic neutrinos can radiatively decay into sterile neutrinos.
This can be probed with 21 cm cosmology and, from EDGES results, we derived constraints  on 
the mass and lifetime of the decaying active neutrino and on the difference of masses between
active and sterile neutrino in the quasi-degenerate case. Interestingly, the scenario can explain the EDGES anomaly 
if this will be confirmed.   The scenario could also potentially have  other testable phenomenological effects such as
the  excess at higher radio frequencies claimed by the ARCADE 2 collaboration. Our results can be also straightforwardly extended
to the case of decaying quasi-degenerate dark matter. Additional independent 
results on the global 21 cm signal from experiments such as SARAS \cite{SARAS} and LEDA \cite{LEDA} 
might provide independent tests of the EDGES anomaly.
If this will be confirmed, a precise determination of the dependence of the absorption signal on redshift could 
potentially be used to test even more  strongly our proposed scenario. 
Certainly 21 cm cosmology opens new fascinating opportunities to test models of new physics and might in a not too far future
finally provide evidence of non-standard cosmological effects. 

\vspace{-1mm}
\subsection*{Acknowledgments}

PDB and MC acknowledge financial support from the STFC Consolidated Grant L000296/1. 
RS is supported  by a Newton International Fellowship from Royal Society (UK) and SERB (India). 
KF acknowledges financial support from the NExT/SEPnet Institute. 
This project has received funding/support from the European Union Horizon 2020 research and innovation 
programme under the Marie Sk\l{}odowska-Curie grant agreement number 690575.
We wish to thank Teppei Katori for useful comments.

\end{document}